\documentclass{aa}      
\usepackage{graphicx}
\usepackage{natbib}
\bibpunct{(}{)}{;}{a}{}{,} 

\begin{document}

\title{Aspherical galaxy clusters: effects on cluster masses and gas mass fractions}

\author{R. Piffaretti\inst{1,2},
        Ph. Jetzer\inst{1} \&
        S. Schindler\inst{3,4}}
\offprints{R. Piffaretti, \\
    \email{piff@physik.unizh.ch}}
\institute{Institute of Theoretical Physics, University of Z\"{u}rich, Winterthurerstrasse, 190, CH-8057 Z\"{u}rich, Switzerland
           \and Paul Scherrer Institute, Laboratory for Astrophysics, CH-5232 Villigen, Switzerland
           \and Institut f\"{u}r Astrophysik, Leopold-Franzens-Universit\"{a}t Innsbruck, Technikerstrasse 25, A-6020 Innsbruck, Austria
          \and Astrophysics Research Institute, Liverpool John Moores University, Twelve Quays House, Birkenhead CH41 1LD, UK    }  
\date{Received  / Accepted}

\abstract{We present an investigation of the effects of asphericity on the estimates of total mass and gas mass fraction in galaxy clusters from X-ray observations. We model the aspherical shape of galaxy clusters by a triaxial model and compare the true total mass and the true total gas mass fraction with the corresponding quantities obtained with the assumption of spherical symmetry. In the triaxial model we allow the extent along the line of sight to vary in order to describe elongated and compressed cluster shapes. Using a sample of 10 ROSAT clusters and a recent CHANDRA observation we find the following results. For prolate or oblate shapes the difference between triaxial and spherical model both in the mass and in the gas mass fraction are negligible (less than 3\%). For more aspherical shapes the total mass is underestimated (overestimated) in the centre, if the cluster is compressed (elongated). The gas mass fraction is overestimated for compressed clusters and slightly underestimated for elongated clusters. Comparing X-ray masses with gravitational lensing estimates, we find that elongations along the line of sight can resolve discrepancies of masses determined by the two different methods of up to $\sim 30 \%$. The combination of Sunyaev-Zel'dovich and X-ray observations is useful to measure the elongation of the cluster along the line of sight. As an application, we estimate the elongation of the cluster CL0016+16 with two different approaches, Sunyaev-Zel'dovich measurements and comparison of weak lensing and X-ray masses, and find reasonable agreement.

}
\titlerunning{Aspherical galaxy clusters}
\maketitle


%
\section{Introduction}
\label{Introduction}
Clusters of galaxies are the largest gravitationally bound aggregates of matter in the universe and, for many applications, they can be regarded as being representative for the universe as a whole. In particular the ratio of baryonic to total matter in clusters can be assumed to be representative of the universe, because of the large volume and the fact that clusters are closed systems. Combined with big bang nucleosynthesis calculations and observed light-element abundances, this ratio can be used to constrain the cosmological density parameter $\Omega_\mathrm{m}$. Optical velocity dispersion measurements and gravitational lensing are independent methods used to constrain the total mass in galaxy clusters while X-ray observations and recently the Sunyaev-Zel'dovich (SZ) effect (Grego et al. 2001) are used to estimate not only total masses but also gas mass fractions in clusters. The X-ray method is based on the assumptions that the X-ray emitting gas is an ideal gas in hydrostatic equilibrium and the total mass and gas mass fraction are estimated through the X-ray surface brightness distribution and the gas temperature. In this context, the $\beta$-model (Cavaliere \& Fusco-Femiano 1976) is widely used in X-ray astronomy to parametrise the gas density profile in clusters of galaxies by fitting their surface brightness. It is usually used under the assumption of spherical symmetry. If the images show that the cluster emission is smooth and the isophotes show an elliptical shape, the modelling is improved when an elliptical $\beta$-model is used: Fabricant et al. (1984) studied the non-spherical shape of A2256, McMillan et al. (1989) the morphology of 49 Abell clusters and the elliptical shape of Cl0016+16 was analysed by Neumann \& B\"ohringer (1997) and Hughes \& Birkinshaw (1998). Buote $\&$ Canizares (1996) studied the elliptical shapes of 5 Abell clusters concluding that, for oblate and prolate shapes, the estimates of the gas masses are insensitive to the ellipticities of the X-ray isophotes.\\
The aim of this paper is to investigate to which degree the assumption of spherical symmetry in the estimate of total mass and gas mass fraction in galaxy clusters showing elliptical X-ray emission is accurate. We select 10 galaxy clusters showing a smooth, elliptical emission in the ROSAT archive and we model the intracluster medium (ICM) with a triaxial $\beta$-model and compare the estimates with those of a spherical $\beta$-model. In particular, a triaxial model allows the description of observations of clusters which may be \textit{compressed} or \textit{elongated} along the line of sight, with the latter being probably more frequent, since elongated clusters are selected preferentially due to their higher contrast over the background emission. Since our investigation is focused on a geometrical aspect common to many clusters, we do not present an exhaustive discussion on the results for each single cluster. Instead, we point out the general features common to each sample cluster. In Sect. \ref{model} we describe our model. In Sect. \ref{Sample, morphology and surface brightness fits} we present the sample and the determination of the morphology and the relevant best fit parameters necessary for the estimate of total masses and gas mass fractions, which are discussed in Sect. \ref{Total mass} and \ref{Gas mass fraction}, respectively.
We discuss the influence of asphericity on the comparison of X-ray and weak lensing mass estimates in Sect. \ref{weaklensing} and investigate further possible constrains on the cluster shape by means of the SZ effect in Sect. \ref{SZ}. We present a summary and conclusions in Sect. \ref{conclusions}.\\
If not stated explicitly, $H_0=50 \, \mathrm{km \, s^{-1} \, Mpc^{-1}}$ and $q_\mathrm{0} =0.5$ are assumed. $1\sigma$ uncertainties for the estimated parameters are used throughout this paper.\\

%
\section{The model}
\label{model}
We extend the classical, spherical $\beta$-model to a more general triaxial, ellipsoidal $\beta$-model:
\begin{equation}\label{ne}
n_\mathrm{e}(x, y, z)=n_{\mathrm{e} \, \mathrm{0}} \Bigg( 1 + \frac{x^2}{a^2}+\frac{y^2}{b^2}+\frac{z^2}{c^2}\Bigg)^{-3 \beta /2 },
\end{equation}
for the electron density in the ICM, where $a$, $b$ and $c$ are the core radii in the $x$, $y$ and $z$ directions respectively. Assuming isothermality of the ICM, the X-ray surface brightness, $S_\mathrm{X}$, is proportional to the integral of the square of the electron density along the line of sight. Taking the latter to be along the $z$ axis we obtain:
\begin{equation}\label{Sx}
S_\mathrm{X}(x,y) \propto c \,  n_{\mathrm{e} \, \mathrm{0}}^2 \times \Bigg(  1 + \frac{x^2}{a^2}+\frac{y^2}{b^2}\Bigg)^{-3 \beta +1/2}.
\end{equation}
Thus, the core radii $a$ and $b$ can be determined directly from X-ray images, while some assumptions on the core radius along the line of sight $c$ must be made. On the other hand, if also a possible inclination of the principal axes of the isodensity surfaces described by this triaxial model is included, only a combination of the three core radii can be determined from the X-ray images. The investigation of this possibility, besides being involved, does not substantially improve the modelling and we do not expect that it will change our conclusions. We thus model the X-ray emission with a spherical model and a triaxial model with no inclination angle to the line of sight. For the spherical model $a=b=c\equiv r_\mathrm{c}^{\mathrm{circ}}$ in Eq. (\ref{Sx}) and for the triaxial model we choose the major core radius  $a \equiv r_\mathrm{c}^{\mathrm{ell}}$, the minor core radius $b \equiv e \times r_\mathrm{c}^{\mathrm{ell}}$ and the core radius along the line of sight  $c \equiv i \times r_\mathrm{c}^{\mathrm{ell}}$. As noticed by Grego et al. (2000) for A 370, the total mass density computed under the assumption of hydrostatic equilibrium and isothermality of the intracluster gas can be unphysical for very elliptical gas density distributions. Consequently, only a specific range for the parameter $i$, which depends on the cluster properties, is allowed in the modelling of ellipsoidal ICM distributions with a  triaxial, ellipsoidal $\beta$-model.
%
\section{Sample and morphological analysis}
\label{Sample, morphology and surface brightness fits}
The clusters are selected from the ROSAT archive. Since our goals require a sample of clusters that are regular and show a smooth, elliptical surface brightness, we exclude all the clusters with bimodal or strongly irregular morphology and those with a high spherical symmetry. In order to have a robust estimate of the parameters that constrain the morphology, we select only those clusters for which both ROSAT High Resolution Imager (HRI) and Position Sensitive Proportional Counter (PSPC) observations are available. Because of the lack of detailed temperature maps and for simplicity, we assume that the gas in each cluster is isothermal. The 10 selected clusters with the additional information needed to deproject the surface brightness are listed in Table \ref{bpp}. 
\begin{table}
\begin{center}
\centering
\caption{The values for the basic physical parameters compiled from the literature: the redshift $z$ and the gas temperature $T_{\mathrm{gas}}$. References:[1] Hughes \& Birkinshaw (1998), [2] Matsumoto et al. (2000), [3] Allen \& Fabian (1998), [4] Markevitch et al. (1998), [5] Ettori \& Fabian (1999).}
\label{bpp}
\begin{tabular}{ccc}
\hline
\hline
$\mathrm{Cluster}$&$z$&$T_{\mathrm{gas}}$\\
$\mathrm{}$&$\mathrm{}$&$\mathrm{(keV)}$\\
\hline
Cl0016+16&$0.5455$&$7.55^{+0.72}_{-0.58}[1]$\\
A 478&$0.0881$&$6.40^{+0.25}_{-0.25}[2]$\\
A 1795&$0.0631$&$5.68^{+0.11}_{-0.11}[2]$\\
A 1068&$0.1386$&$5.5^{+1.4}_{-0.9}[3]$\\
A 1413&$0.1427$&$8.5^{+1.3}_{-0.8}[3]$\\
A 2390&$0.231$&$11.1^{+1.0}_{-1.0}[5]$\\
A 2199&$0.0298$&$4.22^{+0.06}_{-0.06}[2]$\\
A 2029&$0.0765$&$8.47^{+0.41}_{-0.36}[3]$\\
A 2597&$0.0852$&$3.6^{+0.2}_{-0.2}[4]$\\
Hydra A&$0.0522$&$3.71^{+0.14}_{-0.14}[2]$\\
\hline
\end{tabular}
\end{center}
\end{table}
For both HRI and PSPC, we prepare exposure corrected images and remove point sources embedded in the cluster emission to prevent contamination on measurements of the ellipticity of the surface brightness. The determination of the isophotes ellipticity and position angle is performed with the MIDAS routine FIT/ELL3, an iterative least-squares method in which the isophotes are free to translate position, and change ellipticity and orientation (Bender \& M\"ollenhof 1987). From each image processed with this algorithm we obtain a set of ellipses with given position angles, half minor and major axes and centre coordinates. Even though our model implies that the isophotes are similar and concentric, such a set of parameters allows us to identify the isophotes which are suitable to describe the shape of the cluster emission. In the PSPC data we notice that for all the clusters the minor to major axis ratio $e$ tends to unity as we approach the centre. We identify this effect as a distortion due to the point spread function (PSF) of the PSPC and thus discard the isophotes within the central $\sim 25$ arcseconds region.
The X-ray emission can be traced up to $R_{\mathrm{out}}$, which is estimated from the surface brightness profile and listed in Table 5. For each cluster we also can identify a distance from the centre beyond which the isophote parameters jump to arbitrary values. As this distance is approximately equal to $R_{\mathrm{out}}$, we also discard the isophotes beyond it.
We then compute the mean values of the minor to major axis ratio $e$, position angle and centre position from the parameters of the remaining isophotes. For the HRI images we also notice the same central feature, but due to the smaller field of view, we are not able to estimate the ellipses parameters for the whole cluster emission. In any case we find that the mean values for the minor to major axis ratio $e$, the position angle and centre computed from HRI and PSPC images agree within the errors when evaluated over the same radial range. For the analysis we use the PSPC images because of the PSPC's superior sensitivity and larger field of view. The relevant morphological parameters (the position angle and the minor to major axis ratio $e$) from the image analysis are listed in Table \ref{PAe}.
\begin{table}
\begin{center}
\centering
\caption{The estimated morphological parameters: the position angle (N over E) and the minor to major axis ratio $e$ and the ranges of allowed elongations ($i_{\mathrm{min}} \leq i \leq i_{\mathrm{max}}$).}
\label{PAe}
\begin{tabular}{ccccc}
\hline
\hline
$\mathrm{Cluster}$&$\mathrm{Position}$&$e$&$i_{\mathrm{min}}$&$i_{\mathrm{max}}$\\
$\mathrm{}$&$\mathrm{angle (deg)}$&$\mathrm{}$&$\mathrm{}$&$\mathrm{}$\\
\hline
Cl0016+16&$47.1\pm7.7$&$0.85 \pm 0.06$&$0.64$&$1.80$\\ 
A 478&$47.4\pm7.4$&$0.84 \pm 0.06$&$0.65$&$1.56$\\ 
A 1795&$10.2\pm6.4$&$0.83 \pm 0.06$&$0.64$&$1.48$\\ 
A 1068&$-47.0\pm4.9$&$0.79 \pm 0.04$&$0.62$&$1.27$\\ 
A 1413&$-5.1\pm4.3$&$0.72 \pm 0.03$&$0.59$&$1.05$\\  
A 2390&$-60.0\pm8.1$&$0.79 \pm 0.07$&$0.62$&$1.31$\\ 
A 2199&$32.9\pm9.2$&$0.86 \pm 0.04$&$0.66$&$1.68$\\  
A 2029&$16.1\pm7.4$&$0.80 \pm 0.06$&$0.63$&$1.35$\\  
A 2597&$-35.3\pm5.1$&$0.79 \pm 0.06$&$0.62$&$1.30$\\  
Hydra A&$-32.9\pm7.1$&$0.87 \pm 0.05$&$0.66$&$1.75$\\  
\hline
\end{tabular}
\end{center}
\end{table}
Since we are interested in the differences between spherical and ellipsoidal geometries in the estimate of total masses and gas mass fractions, we extract two surface brightness profiles from the PSPC images, referred to as circular and elliptical profiles. The circular profiles are determined by computing the counts within circular annuli spaced by 5 arcsec, while for the elliptical profiles the bins are similar and concentric ellipses with the minor to major axis ratios $e$ and position angles listed in Table 2 and  spaced by 5 arcsec in the direction of the major axis. We parametrise the circular and elliptical profiles with:
\begin{equation}\label{Sxfit}
S_\mathrm{X}(r)=S_\mathrm{0}^j \Big(1+\Big(\frac{r}{r_\mathrm{c}^j}\Big)^2\Big)^{-3 \beta^j + 1/2} + BG^j,
\end{equation}
where $j=\mathrm{circ}, \, \mathrm{ell}$, labels the circular and elliptical models, respectively. For the elliptical profile, the distance $r$ from the centre is measured along the major axis and thus the elliptical core radius $r_\mathrm{c}^{\mathrm{ell}}$ lies along the major axis. The best fit parameters are listed in Tables \ref{fpcirc} and \ref{fpell}.
\begin{table}
\begin{center}
\centering
\caption{The best fit parameters for the circular profile: the central surface brightness $S_\mathrm{0}^{\mathrm{circ}}$, the slope $\beta^{\mathrm{circ}}$ and the circular core radius $r_\mathrm{c}^{\mathrm{circ}}$.}
\label{fpcirc}
\begin{tabular}{cccc}
\hline
\hline
$\mathrm{Cluster}$&$S_\mathrm{0}^{\mathrm{circ}}$&$\beta^{\mathrm{circ}}$&$r_\mathrm{c}^{\mathrm{circ}}$\\
$$&$\mathrm{(10^{-5} \frac{PSPC \, cts}{s \, arcsec^{2}})}$&$$&$\mathrm{(arcsec), \, (kpc)}$\\
\hline
Cl0016+16&$0.91 \pm 0.05$&$0.81 \pm 0.03$&$51.51 \pm 2.94, \; 379$\\
A 478&$12.35 \pm 0.36$&$0.60 \pm 0.01$&$39.91 \pm 0.98, \; 88$\\
A 1795&$12.94 \pm 0.24$&$0.60 \pm 0.01$&$51.76 \pm 0.83, \; 85$\\
A 1068&$12.27 \pm 0.73$&$0.68 \pm 0.01$&$25.73 \pm 1.30, \; 83$\\
A 1413&$3.51 \pm 0.20$&$0.66 \pm 0.01$&$53.74 \pm 2.88, \; 176$\\
A 2390&$4.52 \pm 0.29$&$0.65 \pm 0.01$&$41.09 \pm 2.33, \; 192$\\
A 2199&$6.70 \pm 0.14$&$0.57 \pm 0.01$&$84.55 \pm 1.58, \; 70$\\
A 2029&$17.92 \pm 0.46 $&$0.57 \pm 0.01$&$40.66 \pm 0.90, \; 79$\\
A 2597&$16.85 \pm 0.73$&$0.65 \pm 0.01$&$30.04 \pm 1.09, \; 64$\\
Hydra A&$15.15 \pm 0.54$&$0.57 \pm 0.01$&$35.14 \pm 1.00, \; 49$\\
\hline
\end{tabular}
\end{center}
\end{table}
\begin{table}
\begin{center}
\centering
\caption{The best fit parameters for the elliptical profile: the central surface brightness $S_\mathrm{0}^{\mathrm{ell}}$, the slope $\beta^{\mathrm{ell}}$ and the core radius along the major axis $r_\mathrm{c}^{\mathrm{ell}}$.}
\label{fpell}
\begin{tabular}{cccc}
\hline
\hline
$\mathrm{Cluster}$&$S_\mathrm{0}^{\mathrm{ell}}$&$\beta^{\mathrm{ell}}$&$r_\mathrm{c}^{\mathrm{ell}}$\\
$$&$\mathrm{(10^{-5} \frac{PSPC \, cts}{s \, arcsec^{2}})}$&$$&$\mathrm{(arcsec), \, (kpc)}$\\
\hline
Cl0016+16&$0.89 \pm 0.04$&$0.82 \pm 0.03$&$57.11 \pm 3.24, \; 420$\\
A 478&$12.20 \pm 0.34$&$0.59 \pm 0.01$&$43.02 \pm 1.02, \; 95$\\
A 1795&$12.30 \pm 0.25$&$0.60 \pm 0.01$&$58.77 \pm 1.04, \; 97$\\
A 1068&$12.12 \pm 0.68$&$0.68 \pm 0.01$&$28.76 \pm 1.37, \; 92$\\
A 1413&$3.46 \pm 0.23$&$0.64 \pm 0.01$&$58.85 \pm 3.65, \; 193$\\
A 2390&$3.99 \pm 0.25$&$0.65 \pm 0.01$&$48.67 \pm 2.84, \; 227$\\
A 2199&$6.73 \pm 0.14$&$0.57 \pm 0.01$&$89.31 \pm 1.66, \; 74$\\
A 2029&$18.44 \pm 0.53$&$0.56 \pm 0.01$&$42.84 \pm 1.06, \; 84$\\
A 2597&$16.80 \pm 0.81$&$0.64 \pm 0.01$&$32.72 \pm 1.31, \; 70$\\
Hydra A&$15.06 \pm 0.51$&$0.57 \pm 0.01$&$37.89 \pm 1.01, \; 53$\\
\hline
\end{tabular}
\end{center}
\end{table}
We notice that the central surface brightness of the elliptical profile is in general smaller than that of the circular profile and that the values for $\beta$ are almost the same for both models. Obviously the elliptical core radius $r_\mathrm{c}^{\mathrm{ell}}$ is always larger than $r_\mathrm{c}^{\mathrm{circ}}$.\\

\section{Results for the total mass}
\label{Total mass}

\begin{figure}
 \resizebox{\hsize}{!}{\includegraphics{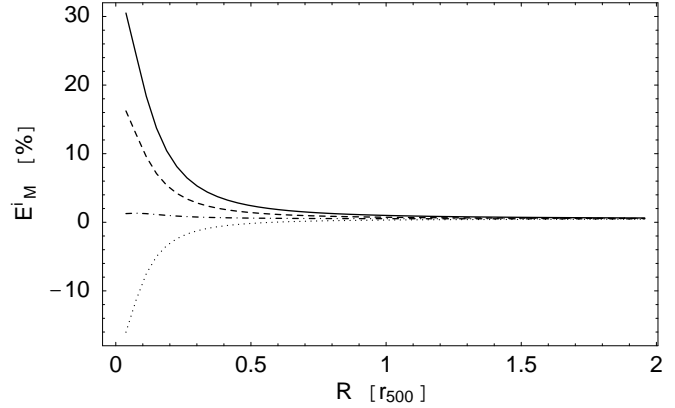}}
 \caption{The relative errors for the total mass estimates for A2390 plotted for the 4 triaxial models: compressed along the line of sight ($i=i_{\mathrm{min}}$) (dotted line), prolate ($i=e$) (dashed-dotted line), oblate ($i=1$) (dashed line) and elongated  ($i=i_{max}$) (solid line) shapes. Positive relative errors imply overestimations of the total mass if spherical symmetry is assumed: this is the case if the cluster is elongated. Underestimations are found for compressed clusters.}
 \label{plotmasses}
\end{figure}

\begin{figure}
 \resizebox{\hsize}{!}{\includegraphics{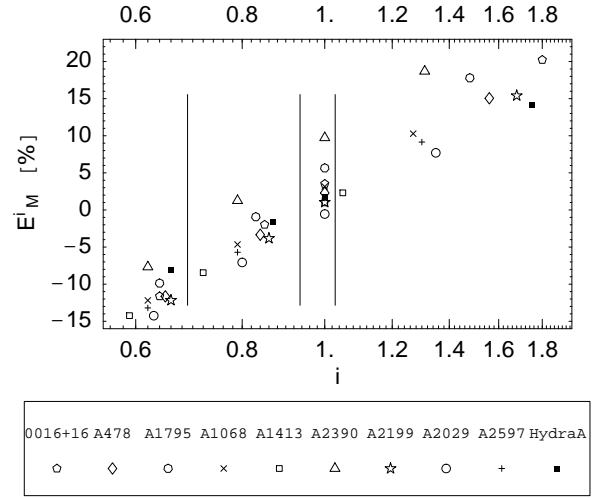}}
 \caption{The relative errors for the total mass $E_\mathrm{M}^i$ as a function of $i$ for the 4 models ($i_{\mathrm{min}}$, $e$, $1$ and $i_{\mathrm{max}}$, which are separated by the vertical lines) at $r_\mathrm{c}^{\mathrm{circ}}$ for the 10 sample clusters. For clusters elongated along the line of sight we find overestimations, while for compressed clusters the total mass is underestimated.}
 \label{EMvsi}
\end{figure}

Assuming hydrostatic equilibrium and isothermality of the intracluster gas, the total mass density $\rho_{\mathrm{tot}}$ of a cluster can be computed from the $\beta$-model with the best fit parameters listed in Tables \ref{fpcirc} and \ref{fpell}:
\begin{equation}\label{rhotot}
\rho_{\mathrm{tot}}=-\Big(\frac{k_\mathrm{B} T_{\mathrm{gas}}}{4\pi G \mu m_\mathrm{p}}\Big)\Delta(ln \rho_{\mathrm{gas}}),
\end{equation}
where $G$ is the gravitational constant, $k_\mathrm{B}$ the Boltzmann constant, $\rho_{\mathrm{gas}}$ the gas mass density, $T_{\mathrm{gas}}$ its temperature and $\mu m_\mathrm{p}$ is the mean particle mass of the gas (we assume $\mu=0.61$). In order to compare the total masses for spherical and ellipsoidal ICM shapes, we compute them by integrating Eq. (\ref{rhotot}) within a sphere of radius $R$. We choose to integrate the total density within a sphere for both models because, besides being the natural choice for the spherical modelling, it allows the comparison of masses contained within the same volume. We quantify the difference between the two models by defining the relative error:
\begin{equation}\label{EMi}
E_\mathrm{M}^i(R)=\frac{M^{\mathrm{sph}}_{\mathrm{tot}}(R)-M^{i}_{\mathrm{tot}}(R)}{M^{i}_{\mathrm{tot}}(R)},
\end{equation}
where $M^{\mathrm{sph}}_{\mathrm{tot}}(R)$ and $M^{i}_{\mathrm{tot}}(R)$ are the total masses within the distance $R$ from the centre for the spherical model and a triaxial model with a core radius along the line of sight $c=i \times  r_\mathrm{c}^{\mathrm{ell}}$, respectively. Since the ellipsoidal modelling should be more accurate than the spherical, positive (negative) values of $E_\mathrm{M}^i$ are going to be characterised as over- (under-) estimations. For spherical symmetry total masses evaluated from the best fit parameters of the circular analysis for $R=r_{\mathrm{500}}$ (the radius which encompasses a volume that has a total mass density 500 times the critical density for closure) are listed in Table \ref{r500andmasses}. 
\begin{table}
\begin{center}
\centering
\caption{$r_{\mathrm{500}}$, the outer radius $R_{\mathrm{out}}$ and the total masses and gas mass fractions for the spherical model for $R=r_{\mathrm{500}}$.}
\label{r500andmasses}
\begin{tabular}{ccccc}
\hline
\hline
$\mathrm{Cluster}$&$r_{\mathrm{500}}$&$R_{\mathrm{out}}$&$M_{\mathrm{tot}}^{\mathrm{sph}}(r_{\mathrm{500}})$&$f_{\mathrm{gas}}^{\mathrm{sph}}(r_{\mathrm{500}})$\\
$$&$\mathrm{(Mpc)}$&$(r_{\mathrm{500}})$&$\mathrm{(10^{14} \, M_{\sun})}$&$$\\
\hline
Cl0016+16&$1.051$&$2.1$&$6.23 \pm 0.93$&$0.268 \pm 0.027$\\
A 478&$1.492$&$1.5$&$6.22 \pm 0.34$&$0.256 \pm 0.019$\\
A 1795&$1.456$&$1.2$&$5.39 \pm 0.20$&$0.217 \pm 0.015$\\
A 1068&$1.384$&$1.2$&$5.69 \pm 1.55$&$0.131 \pm 0.018$\\
A 1413&$1.677$&$1.9$&$10.23 \pm 1.73$&$0.157 \pm 0.013$\\
A 2390&$1.731$&$1.9$&$14.05 \pm 2.18$&$0.186 \pm 0.015$\\
A 2199&$1.288$&$1.3$&$3.39 \pm 0.11$&$0.179 \pm 0.013$\\
A 2029&$1.713$&$1.1$&$9.11 \pm 0.61$&$0.212 \pm 0.019$\\
A 2597&$1.172$&$1.4$&$2.99 \pm 0.21$&$0.183 \pm 0.011$\\
Hydra A&$1.170$&$1.3$&$2.71 \pm 0.15$&$0.207 \pm 0.017$\\
\hline
\end{tabular}
\end{center}
\end{table}
In the case of ellipsoidal shapes some constrains on the core radius along the line of sight, $c=i \times r_\mathrm{c}^{\mathrm{ell}}$, must be taken into account.
As outlined in Sect. \ref{model}, the total mass density inferred by means of Eq. (\ref{rhotot}) can be unphysical for certain ellipsoidal shapes of the ICM. In fact, very elliptical gas density distributions would imply regions with negative total mass density. As we model the cluster within a sphere with radius $R_{\mathrm{out}}$, we clearly require the total mass density to be be positive inside such a sphere. Consequently, we obtain a range for the parameter $i$ ($i_{\mathrm{min}} \leq i \leq i_{\mathrm{max}}$) for each sample cluster. These values are listed in Table \ref{PAe}. Notice that very ellipsoidal shapes are not allowed. We emphasise that the computed range $i_{\mathrm{min}} \leq i \leq i_{\mathrm{max}}$ implies negative values for the total density only at distances abundantly larger than $R_{\mathrm{out}}$, where anyhow we do no longer expect the adopted triaxial model to be a reasonable approximation. Indeed, at very large radii the shape of the cluster is dominated by the infall of galaxies and groups of galaxies. Therefore, the cluster shape there has certainly no similarity to the shape of the inner region, and hence the problem of a negative total mass density does not arise. 
We present the results of our investigation of the relative errors $E_\mathrm{M}^i$ for: the two axisymmetric ellipsoidal geometries; oblate ($c=r_\mathrm{c}^{\mathrm{ell}}$, $i=1$ in Eq. (\ref{EMi})) and prolate ($i=e$) spheroids, a shape for which the distribution is the most compressed along the line of sight ($i=i_{\mathrm{min}}$) and a shape that describes the most elongated shape ($i=i_{\mathrm{max}}$). For simplicity, we show in Fig. \ref{plotmasses} the relative errors $E_\mathrm{M}^i(R)$ for these 4 ellipsoidal models only for the cluster A2390, since the trend of these functions is qualitatively the same for all the sample clusters.
We notice that the errors for the total masses in Table 5 include the uncertainties on the gas temperatures, which are quite large. Instead, if this is not included, the errors are of the order of $2-3 \%$. Consequently, errors of this order which are solely due to the non-spherical modelling of the cluster are irrelevant. We thus emphasise that only the relative errors  $E_\mathrm{M}^i(R)$ larger than $\sim 3 \%$ are significant.\\
Since most of the mass is contained in the core, we find that at large radii, at $r_{\mathrm{500}}$ for instance, the relative errors $E_\mathrm{M}^i$ are almost independent on $i$ and have values of a few per cent. This means that at large distances from the centre the possible compression or elongation along the line of sight is not important and that the assumption of spherical symmetry yields very reliable results. The relative errors for $R=r_{\mathrm{500}}$ for the 4 models are listed in Table \ref{EMatr500}.
\begin{table}
\begin{center}
\centering
\caption{The relative error for the total masses for $R=r_{\mathrm{500}}$.}
\label{EMatr500}
\begin{tabular}{ccccc}
\hline
\hline
$\mathrm{Cluster}$&$E_\mathrm{M}^{i_{\mathrm{min}}}$&$E_\mathrm{M}^{e}$&$E_\mathrm{M}^{1}$&$E_\mathrm{M}^{i_{\mathrm{max}}}$\\
$$&$\mathrm{(\%)}$&$\mathrm{(\%)}$&$\mathrm{(\%)}$&$\mathrm{(\%)}$\\
\hline
Cl0016+16&$-3.44$&$-1.36$&$-0.08$&$4.72$\\
A 478&$0.84$&$0.89$&$0.93$&$1.04$\\
A 1795&$-0.65$&$-0.60$&$-0.55$&$-0.45$\\
A 1068&$0.73$&$0.78$&$0.83$&$0.89$\\
A 1413&$3.73$&$3.84$&$4.06$&$4.09$\\
A 2390&$0.32$&$0.52$&$0.73$&$1.00$\\
A 2199&$0.85$&$0.90$&$0.93$&$1.04$\\
A 2029&$2.05$&$2.08$&$2.10$&$2.15$\\
A 2597&$1.51$&$1.55$&$1.59$&$1.65$\\
Hydra A&$-0.20$&$-0.17$&$-0.16$&$-0.08$\\
\hline
\end{tabular}
\end{center}
\end{table}
\begin{table}
\begin{center}
\centering
\caption{The relative error for the total masses for $R=r_{\mathrm{c}}^{\mathrm{circ}}$.}
\label{}
\begin{tabular}{ccccc}
\hline
\hline
$\mathrm{Cluster}$&$E_\mathrm{M}^{i_{\mathrm{min}}}$&$E_\mathrm{M}^{e}$&$E_\mathrm{M}^{1}$&$E_\mathrm{M}^{i_{\mathrm{max}}}$\\
$$&$\mathrm{(\%)}$&$\mathrm{(\%)}$&$\mathrm{(\%)}$&$\mathrm{(\%)}$\\
\hline
Cl0016+16&$-11.62$&$-1.98$&$3.52$&$20.23$\\
A 478&$-11.64$&$-3.34$&$2.29$&$15.05$\\
A 1795&$-9.87$&$-0.93$&$5.65$&$17.79$\\
A 1068&$-12.19$&$-4.65$&$3.16$&$10.28$\\
A 1413&$-14.24$&$-8.45$&$0.96$&$2.31$\\
A 2390&$-7.68$&$1.26$&$9.73$&$18.69$\\
A 2199&$-12.18$&$-3.83$&$1.03$&$15.40$\\
A 2029&$-14.25$&$-7.07$&$-0.55$&$7.70$\\
A 2597&$-13.20$&$-5.70$&$1.61$&$9.15$\\
Hydra A&$-8.13$&$-1.67$&$1.77$&$14.08$\\
\hline
\end{tabular}
\end{center}
\end{table}
As $R$ decreases the relative errors in the mass estimations get larger (see Fig. \ref{plotmasses}): for shapes elongated along the line of sight we always find overestimations which increase with the value of the core radius along the line of sight. The reason is that for strong elongations a large portion of the core is excluded from the sphere within which the mass is estimated. A similar explanation holds for the underestimation found for compressed shapes. In Fig. \ref{EMvsi}, we show the relative errors at $R=r_\mathrm{c}^{\mathrm{circ}}$ for the 10 sample clusters as a function of $i$.
At $R=r_\mathrm{c}^{\mathrm{circ}}$ we find: $<E_\mathrm{M}^{i_{\mathrm{min}}}>_\mathrm{S}=-12  \, \%$, $<E_\mathrm{M}^{e}>_\mathrm{S}=-4  \, \%$, $<E_\mathrm{M}^{1}>_\mathrm{S}=3  \, \%$, $<E_\mathrm{M}^{i_{\mathrm{max}}}>_\mathrm{S}=13  \, \%$, where $<\;>_\mathrm{S}$ means that we average over the 10 sample clusters. As the masses are usually estimated at $R=r_{\mathrm{500}}$ or larger, these errors usually do not affect the mass determination.\\
We also perform the same analysis on a recent CHANDRA observation of the galaxy cluster RBS797 (Schindler et al. 2001), which shows an elliptical emission with a minor to major axis ratio of $0.77$. This cluster is an excellent object for this analysis because the minor to major axis ratio $e$ and the position angle do hardly change with radius. For this cluster we find $M_{\mathrm{tot}}^{\mathrm{sph}}(r_{\mathrm{500}})=6.5 \times 10^{14} \mathrm{M}_{\sun}$ and  $f_{\mathrm{gas}}^{\mathrm{sph}}(r_{\mathrm{500}})=0.17$ and we arrive at the same conclusions on the relative errors for the total mass and gas mass fraction  obtained from our ROSAT sample. For instance, comparing these values with the estimates from the 4 triaxial models discussed in this section, we find that the relative error for the total mass within $r_{\mathrm{500}}$ is always less than $1 \%$.
%
\section{Results for the gas mass fraction}
\label{Gas mass fraction}
\begin{figure}
 \resizebox{\hsize}{!}{\includegraphics{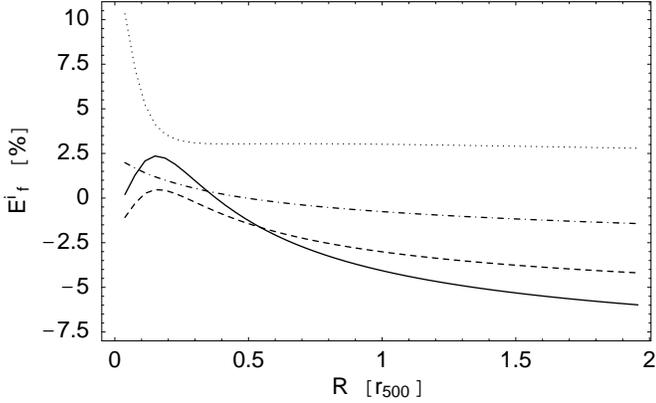}}
 \caption{The relative errors for the gas mass fraction estimates for A2390 plotted for the 4 triaxial models: compressed along the line of sight ($i=i_{\mathrm{min}}$) (dotted line), prolate ($i=e$) (dashed-dotted line), oblate ($i=1$) (dashed line) and elongated  ($i=i_{\mathrm{max}}$) (solid line) shapes. For compressed clusters the gas mass fraction is overestimated.}
 \label{plotfg}
\end{figure}

\begin{figure}
 \resizebox{\hsize}{!}{\includegraphics{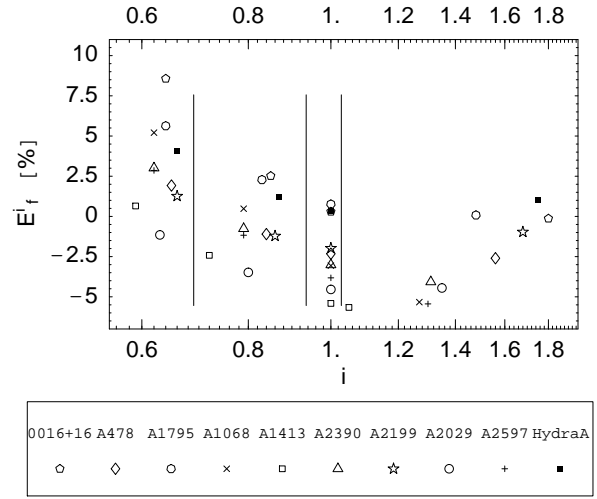}}
 \caption{The relative error for the gas mass fraction $E^i_\mathrm{f}$ as a function of $i$ for the 4 models ($i_{\mathrm{min}}$, $e$, $1$ and $i_{\mathrm{max}}$) at $r_{\mathrm{500}}$ for the 10 sample clusters. The 4 elongations are separated by 3 vertical lines.}
 \label{Efir500}
\end{figure}

We investigate the same 4 triaxial models as in Sect. \ref{Total mass} with respect to the gas mass fraction. For the computation of the gas mass fractions for the 10 sample clusters we deproject the X-ray emission with the $\beta$-model in order to determine the gas density $\rho_{\mathrm{gas}}$ and then integrate it within a sphere of radius $R$. We thus obtain the gas mass fractions $f_{\mathrm{gas}}^{\mathrm{sph}}(R)$ for the spherical geometry, which are listed in Table \ref{r500andmasses} for $R=r_{\mathrm{500}}$, and the gas mass fractions $f_{\mathrm{gas}}^{i}(R)$ for the triaxial models, where $i$ is defined as in Sect. \ref{model}. Similarly, we define the relative error: 
 \begin{equation}\label{Efi}
E_\mathrm{f}^i(R)=\frac{f^{\mathrm{sph}}_{\mathrm{gas}}(R)-f^{i}_{\mathrm{gas}}(R)}{f^{i}_{\mathrm{gas}}(R)}.
\end{equation} 
For the reason discussed in Sect. \ref{Total mass}, only relative errors for the gas mass fractions larger than $\sim 3 \%$ are of significance. The general trend of the relative errors $E_\mathrm{f}^i(R)$ is the same for all the sample clusters and can be seen in Fig. \ref{plotfg}, where the relative errors for the cluster A2390 are shown.\\
Since the gas mass fraction $f^{i}_{\mathrm{gas}}(R)$ is the ratio of the gas mass $M_{\mathrm{gas}}^i(R)$ to the total mass $M_{\mathrm{tot}}^i(R)$, the discussion of the relative errors $E_\mathrm{f}^i(R)$ involves many quantities.  We notice that, since the central surface brightness is proportional to $c \times \rho_{\mathrm{gas} \, \mathrm{0}}^2$, the central gas density is larger for the compressed shapes than for elongated.
This factor is usually dominant for small $R$, leading to gas mass overestimations for elongated ellipsoids and underestimations for compressed shapes.
In Table \ref{Efrc} we list the gas mass fraction relative errors for the 4 models at $R=r_\mathrm{c}^{\mathrm{circ}}$. We find small over- and underestimations for oblate and prolate shapes for small $R$: $<|E_\mathrm{f}^{1}|>_\mathrm{S}=2  \, \%$ and  $<|E_\mathrm{f}^{e}|>_\mathrm{S}=3  \, \%$ at $R=r_\mathrm{c}^{\mathrm{circ}}$. For elongated and compressed shapes we find larger but in general still negligible errors: at $R=r_\mathrm{c}^{\mathrm{circ}}$ we get $<E_\mathrm{f}^{i_{\mathrm{min}}}>_\mathrm{S}=6  \, \%$ and  $<E_\mathrm{f}^{i_{\mathrm{max}}}>_\mathrm{S}=5  \, \%$. \\
\begin{table}
\begin{center}
\centering
\caption{Relative errors for the gas mass fractions at $R=r_\mathrm{c}^{\mathrm{circ}}$.}
\label{Efrc}
\begin{tabular}{ccccc}
\hline
\hline
$\mathrm{Cluster}$&$E_{\mathrm{f}}^{i_{\mathrm{min}}}$&$E_{\mathrm{f}}^{e}$&$E_{\mathrm{f}}^{1}$&$E_{\mathrm{f}}^{i_{\mathrm{max}}}$\\
$$&$\mathrm{(\%)}$&$\mathrm{(\%)}$&$\mathrm{(\%)}$&$\mathrm{(\%)}$\\
\hline
Cl0016+16&$7.54$&$1.37$&$-0.29$&$4.94$\\

A 478&$4.91$&$1.90$&$1.38$&$6.19$\\

A 1795&$3.96$&$0.64$&$-0.04$&$3.70$\\

A 1068&$6.78$&$2.94$&$1.57$&$2.94$\\

A 1413&$10.74$&$7.73$&$6.57$&$6.81$\\

A 2390&$5.31$&$1.43$&$0.24$&$2.05$\\

A 2199&$4.53$&$1.83$&$1.52$&$8.06$\\

A 2029&$6.40$&$3.88$&$3.56$&$6.48$\\

A 2597&$7.44$&$3.93$&$2.94$&$4.82$\\

Hydra A&$2.30$&$0.40$&$0.40$&$7.92$\\
\hline
\end{tabular}
\end{center}
\end{table}
At large distances the core radius along the line of sight is dominant: small core radii imply steep profiles and consequently less gas mass. In Fig. \ref{Efir500}, we show the relative errors at $R=r_{\mathrm{500}}$ for the 10 sample clusters as a function of $i$. We see that at large distances from the cluster center, at $r_{\mathrm{500}}$ or larger, where the total masses for a spherical and an elliptical model are almost the same (see Sect. \ref{Total mass}), the compressed shapes imply an overestimation of the gas mass fraction for the majority of the clusters. Although this overestimation is small, the more compressed the shape, the larger the overestimation (see Fig. \ref{plotfg}). 
At $R=r_{\mathrm{500}}$ we find the average values: $<E_\mathrm{f}^{i_{\mathrm{min}}}>_\mathrm{S}=3  \, \%$, and $<E_\mathrm{f}^{i_{\mathrm{max}}}>_\mathrm{S}=-2.8  \, \%$ at $R=r_{\mathrm{500}}$, showing small underestimations of the gas mass fractions for elongated shapes (see also Table \ref{Efr500}). For oblate ($i=1$) and prolate ($i=e$) shapes we find that at $R=r_{\mathrm{500}}$ both underestimates and overestimates are present and that they are small (see Table \ref{Efr500}).
\begin{table}
\begin{center}
\centering
\caption{Relative errors for the gas mass fractions at $R=r_{\mathrm{500}}$.}
\label{Efr500}
\begin{tabular}{ccccc}
\hline
\hline
$\mathrm{Cluster}$&$E_{\mathrm{f}}^{i_{\mathrm{min}}}$&$E_{\mathrm{f}}^{e}$&$E_{\mathrm{f}}^{1}$&$E_{\mathrm{f}}^{i_{\mathrm{max}}}$\\
$$&$\mathrm{(\%)}$&$\mathrm{(\%)}$&$\mathrm{(\%)}$&$\mathrm{(\%)}$\\
\hline
Cl0016+16&$8.57$&$2.52$&$0.30$&$-0.13$\\

A 478&$1.92$&$-1.10$&$-2.32$&$-2.60$\\

A 1795&$5.63$&$2.28$&$0.76$&$0.08$\\

A 1068&$5.21$&$0.48$&$-3.06$&$-5.33$\\

A 1413&$0.65$&$-2.42$&$-5.41$&$-5.66$\\

A 2390&$3.01$&$-0.76$&$-3.01$&$-4.07$\\

A 2199&$1.26$&$-1.22$&$-1.98$&$-0.97$\\

A 2029&$-1.15$&$-3.48$&$-4.54$&$-4.45$\\

A 2597&$2.86$&$-1.17$&$-3.82$&$-5.44$\\

Hydra A&$4.07$&$1.19$&$0.34$&$1.05$\\
\hline
\end{tabular}
\end{center}
\end{table}
Averaging the absolute values of the relative errors we find: $<|E_\mathrm{f}^{e}|>_\mathrm{S}=1.7  \, \%$ and  $<|E_\mathrm{f}^{1}|>_\mathrm{S}=2.6 \, \%$. We conclude that these errors usually do not substantially affect the gas mass fraction determination at large radii.\\
%
\section{Asphericity and comparisons with gravitational lensing}
\label{weaklensing}
\begin{figure}
 \resizebox{\hsize}{!}{\includegraphics{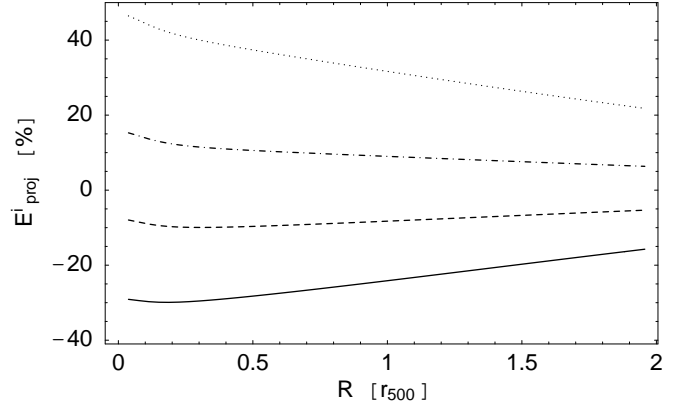}}
 \caption{The relative errors $E^i_{\mathrm{proj}}(R)$ for the \textit{projected} mass estimates for A2390 plotted for the 4 triaxial models: compressed along the line of sight ($i=i_{\mathrm{min}}$) (dotted line), prolate ($i=e$) (dashed-dotted line), oblate ($i=1$) (dashed line) and elongated  ($i=i_{\mathrm{max}}$) (solid line) shapes. The cut-off is $4 \times r_{\mathrm{500}}$. Negative relative errors imply underestimations of the total mass if spherical symmetry is assumed: this is the case if the cluster is elongated. Overestimations are found for compressed clusters. Therefore, an elongation along the line of sight can contribute to resolve the discrepancy between X-ray and lensing mass.}
 \label{Eprojplot}
\end{figure}

Since the X-ray method and gravitational lensing are independent methods used to determine the dark matter distribution in cluster of galaxies, their comparison is very useful. First of all, the morphology of the dark matter distribution can be compared: Smail et al. (1995) found good agreement for the position angle and ellipticity of Cl0016+16 determined with weak lensing and X-ray observations. Furthermore, Miralda-Escud\'{e} \& Babul (1995), presenting a detailed study of three clusters with both arcs and X-ray data available, conclude that the mass estimates from the arc modelling can be nearly a factor $\sim \, 2-3$ larger than those from the X-ray observation in the innermost regions. As summarised by Kneib (2000), the discrepancy might be due to different reasons: too simple X-ray modelling, merging and projection effects. Also non-thermal effects can play a role, although magnetic fields can be ruled out as error sources (Dolag \& Schindler 2000).\\
Since gravitational lensing estimates provide \textit{projected} masses within \textit{projected} distances from the centre, the X-ray mass must be projected too, in order to achieve the comparison. In Sect. \ref{Total mass} the X-ray mass was calculated within a spherical volume, while for the comparison it has to be computed within a cylindrical volume. We investigate projection effects by considering a triaxial $\beta$-model and the 4 models used in Sect. \ref{Total mass} and \ref{Gas mass fraction}. As done in Sect. \ref{Total mass}, we compare the estimates from the triaxial modelling with those obtained assuming spherical symmetry, but integrating Eq. (\ref{rhotot}) within a cylinder of radius $R$. We thus obtain a projected masses $M^{\mathrm{sph}}_{\mathrm{proj}}(R)$ and $M^{i}_{\mathrm{proj}}(R)$  for the spherical model and a triaxial model with a core radius along the line of sight $c=i \times  r_\mathrm{c}^{\mathrm{ell}}$, respectively. Formally, the integral of the total mass density extends from the observer along the line of sight through the cluster infinitely; in practice, a cutoff is used. In the following, a cutoff equal to $4 \times r_{\mathrm{500}}$ is used, as a compromise between to much extrapolation and too short integration length. Although the integration volume is different from the one used in section \ref{Total mass}, the ranges of elongations $i$ given in Table \ref{PAe} apply in this case as well.\\
Similar to the relative errors defined in Sect. \ref{Total mass} and \ref{Gas mass fraction}, we define: 
 \begin{equation}\label{EMiproj}
E_{\mathrm{proj}}^i(R)=\frac{M^{\mathrm{sph}}_{\mathrm{proj}}(R)-M^{i}_{\mathrm{proj}}(R)}{M^{i}_{\mathrm{proj}}(R)}.
\end{equation}     
Since mass estimates from the arc modelling are usually larger than those from the X-ray observations, a triaxial model that implies a negative relative error can alleviate the discrepancy between lensing and X-ray mass estimates. This is the case if the triaxial model describes elongated shapes, as shown in Fig. \ref{Eprojplot}, where we plot the relative errors $E_{\mathrm{proj}}^i(R)$ for the cluster A2390 (the trend of the relative errors is the same for all the sample clusters).\\ 
In Table \ref{Eprojrc} we list the relative errors at $R=r_\mathrm{c}^{\mathrm{circ}}$ for $i=i_{\mathrm{min}},e,1$ and $i_{\mathrm{max}}$. Within this projected distance we find the average values: $<E^{i_{\mathrm{min}}}_{\mathrm{proj}}>_\mathrm{S}=40  \, \%$, $<E^{e}_{\mathrm{proj}}>_\mathrm{S}=10  \, \%$, $<E^{1}_{\mathrm{proj}}>_\mathrm{S}=-10  \, \%$ and  $<E^{i_{\mathrm{max}}}_{\mathrm{proj}}>_\mathrm{S}=-36  \, \%$.   
\begin{table}
\begin{center}
\centering
\caption{Relative errors for the \textit{projected} masses at $R=r_\mathrm{c}^{\mathrm{circ}}$ for the 4 models. The cutoff along the line of sight is $4 \times r_{\mathrm{500}}$.}
\label{Eprojrc}
\begin{tabular}{ccccc}
\hline
\hline
$\mathrm{Cluster}$&$E_{\mathrm{proj}}^{i_{\mathrm{min}}}$&$E_{\mathrm{proj}}^{e}$&$E_{\mathrm{proj}}^{1}$&$E_{\mathrm{proj}}^{i_{\mathrm{max}}}$\\
$$&$\mathrm{(\%)}$&$\mathrm{(\%)}$&$\mathrm{(\%)}$&$\mathrm{(\%)}$\\
\hline
Cl0016+16&$41.54$&$11.03$&$-3.55$&$-41.07$\\

A 478&$39.43$&$8.29$&$-8.61$&$-40.80$\\

A 1795&$41.24$&$9.71$&$-8.81$&$-37.81$\\

A 1068&$40.01$&$11.08$&$-12.38$&$-30.63$\\

A 1413&$39.10$&$14.26$&$-16.44$&$-20.27$\\

A 2390&$43.68$&$13.70$&$-9.01$&$-29.71$\\

A 2199&$38.49$&$6.89$&$-7.93$&$-44.57$\\

A 2029&$38.75$&$9.11$&$-11.98$&$-34.46$\\

A 2597&$40.26$&$10.47$&$-12.24$&$-32.12$\\

Hydra A&$39.68$&$6.67$&$-7.19$&$-46.37$\\
\hline
\end{tabular}
\end{center}
\end{table}
In Table \ref{Eprojr500} we list the relative errors at $R=r_{\mathrm{500}}$ for the 4 models. Within  $R=r_{\mathrm{500}}$ we find the average values: $<E^{i_{\mathrm{min}}}_{\mathrm{proj}}>_\mathrm{S}=33  \, \%$, $<E^{e}_{\mathrm{proj}}>_\mathrm{S}=9  \, \%$, $<E^{1}_{\mathrm{proj}}>_\mathrm{S}=-6  \, \%$ and  $<E^{i_{\mathrm{max}}}_{\mathrm{proj}}>_\mathrm{S}=-27  \, \%$.   
\begin{table}
\begin{center}
\centering
\caption{Relative errors for the \textit{projected} masses at $R=r_{\mathrm{500}}$ for the 4 models. The cutoff along the line of sight is $4 \times r_{\mathrm{500}}$.}
\label{Eprojr500}
\begin{tabular}{ccccc}
\hline
\hline
$\mathrm{Cluster}$&$E_{\mathrm{proj}}^{i_{\mathrm{min}}}$&$E_{\mathrm{proj}}^{e}$&$E_{\mathrm{proj}}^{1}$&$E_{\mathrm{proj}}^{i_{\mathrm{max}}}$\\
$$&$\mathrm{(\%)}$&$\mathrm{(\%)}$&$\mathrm{(\%)}$&$\mathrm{(\%)}$\\
\hline
Cl0016+16&$40.70$&$14.35$&$1.65$&$-30.89$\\

A 478&$30.77$&$7.31$&$-5.70$&$-30.72$\\

A 1795&$29.50$&$6.27$&$-7.69$&$-29.78$\\

A 1068&$31.44$&$9.65$&$-8.45$&$-22.69$\\

A 1413&$35.65$&$16.04$&$-8.78$&$-11.90$\\

A 2390&$31.33$&$9.01$&$-8.27$&$-24.14$\\

A 2199&$30.42$&$6.55$&$-4.88$&$-33.42$\\

A 2029&$32.91$&$10.20$&$-6.35$&$-24.18$\\

A 2597&$32.84$&$10.22$&$-7.45$&$-23.10$\\ 

Hydra A&$29.60$&$5.04$&$-5.51$&$-35.57$\\
\hline
\end{tabular}
\end{center}
\end{table}
These results show that X-ray estimated projected masses are larger for elongated clusters than for spherical ones and that discrepancies up to $\sim 30 \%$ between the two methods can be resolved if a triaxial $\beta$-model with a maximal elongation $i_{\mathrm{max}}$ is used.\\
For the cluster Cl0016+16, Smail et al. (1995) derive a \textit{projected} mass of $7.3 \times 10^{14} \mathrm{M}_{\sun}$ integrated out to a projected radius of $600 \, \mathrm{kpc}$ using X-rays, and  $8.5 \times 10^{14} \mathrm{M}_{\sun}$ from weak lensing. Using the spherical model, our estimate for the projected mass is  $6 \times 10^{14} \mathrm{M}_{\sun}$ if no cut-off is used and  $5.6 \times 10^{14} \mathrm{M}_{\sun}$ with a cutoff equal to $4 \times r_{\mathrm{500}}$. Instead we find that using a triaxial model the discrepancy disappears if the core radius along the line of sight is $\sim 1.76 \times r_\mathrm{c}^{\mathrm{ell}}$ (we use a cutoff equal to $4 \times r_{\mathrm{500}}$). Cl0016+16 is among the most X-ray luminous clusters known and X-ray selected, thus a shape which is elongated along the line of sight is not surprising. From this estimate we conclude that elongation is one of the factors contributing to the discrepancy between lensing and X-ray mass estimates.

\section{Constraints from the Sunyaev-Zel'dovich effect}
\label{SZ}
From X-ray observations only, it is impossible to have information on the cluster elongation in the direction of the line of sight. The uncertainties on the total mass and gas mass fraction estimates due to this unknown quantity might be reduced by noticing that the core radius along the line of sight can be constrained using a complementary and independent measurement: the SZ effect. Assuming the triaxial $\beta$-model as in Sect. (\ref{Introduction}) one gets:
\begin{equation}\label{SZform}
i \propto \frac{\tilde{y}^2(x,y)}{S_\mathrm{X}(x,y)}\frac{H_\mathrm{0}}{r_\mathrm{c}^{\mathrm{ell}}}\times \Bigg(  1 + \Big(\frac{x}{r_\mathrm{c}^{\mathrm{ell}}}\Big)^2+\Big(\frac{y}{e \times r_\mathrm{c}^{\mathrm{ell}}}\Big)^2\Bigg)^{-1/2},
\end{equation}
where $\tilde{y}$ is the SZ Compton parameter, $H_\mathrm{0}$ is the Hubble constant, $r_\mathrm{c}^{\mathrm{ell}}$ is the angular major core radius and $i$ is defined as in Sect. (\ref{Total mass}) (see Puy et al. (2000) for the computation of $S_\mathrm{X}$ and $\tilde{y}$). Thus, the combination of X-ray emission and the SZ intensity change, which is usually used to constrain the Hubble constant under some geometrical assumptions (spherical, oblate or prolate shapes), can be used to estimate the elongation along the line of sight $i$, provided that a value for the Hubble constant is assumed. Although the SZ effect is strongly affected by temperature gradients, we again assume isothermality of the ICM because of the lack of detailed temperature maps. As an example we estimate $i$ for the cluster Cl0016+16 using the measurement by Hughes \& Birkinshaw (1998). Assuming $H_0=50 \, \mathrm{km} \, \mathrm{s}^{-1} \, \mathrm{Mpc}^{-1}$ and the value $\tilde{y}(0,0)=(2.27 \pm 0.36) \times 10^{-4}$, for the Compton parameter measured for the line of sight going through the cluster centre we obtain: $i=0.87^{+0.27}_{-0.22}$. For $i=0.87$ the errors due the assumption of spherical symmetry are negligible, since in this case the relative errors defined in Sect. (\ref{Total mass}) and (\ref{Gas mass fraction}) are less than $3 \, \%$ for any $R$ considered. Taking the errors on the estimated $i$ into account, we find the following results. At $r_{\mathrm{500}}$, the relative errors for the total mass are less than $3 \%$ and those for the gas mass fraction less than $7 \%$. At $r_c^{circ}$, the relative errors for the total mass are $8 \%$ and  $-11 \%$ (for upper and lower limit of $i$, respectively) and, those for the gas mass fraction less than $7 \%$.\\
Instead, using $H_0=71 \, \pm 7\, \mathrm{km} \, \mathrm{s}^{-1} \, \mathrm{Mpc}^{-1}$ (Fukugita \& Hogan 2000), we obtain: $i=1.24^{+0.54}_{-0.40}$. We find: $E^{1.24}_\mathrm{M}(r_{\mathrm{500}})=1.7 \, \%$, $E^{1.24}_\mathrm{f}(r_{\mathrm{500}})=-1.3 \, \%$,  $E^{1.24}_\mathrm{M}(r_\mathrm{c}^{\mathrm{circ}})=10 \, \%$ and $E^{1.24}_\mathrm{f}(r_\mathrm{c}^{\mathrm{circ}})=-0.4 \, \%$. We conclude that in this case a triaxial $\beta$-model improves the estimation of the total mass within the central region. We note that the estimated  elongation of $i \sim 1.24$ agrees qualitatively with the conclusion of Sect. \ref{weaklensing}, but this smaller value suggests that elongations along the line of sight can only partially resolve the discrepancy between lensing and X-ray mass estimates.\\ 

\section{Conclusions}
\label{conclusions}
The elliptical X-ray emission shown by many galaxy clusters make us ponder about the possible triaxial distribution of the ICM. In this paper we model the X-ray surface brightness by means of a spherical and a triaxial $\beta$-model and, assuming hydrostatic equilibrium and isothermality of the intracluster gas, we compare the estimates on the total mass and gas mass fraction we obtain from these geometrically different models. Strongly elongated shapes are not allowed in this model, suggesting that if these should be nonetheless observed they may be indicative of unvirialised clusters. Analysing 10 ROSAT clusters we find:
\begin{itemize}
\item If prolate or oblate shapes are assumed, then the differences in the estimates of both total mass and gas mass fraction are negligible (less than $3 \%$) at every distance from the cluster centre.
\item If the cluster is compressed along the line of sight, the total mass in the central regions of the cluster is underestimated. The underestimation depends on the degree of compression, the more compressed the shape, the larger the underestimation.
\item  If the cluster is elongated along the line of sight, the total mass in the central regions is overestimated. The overestimation depends on the elongation: the more elongated the ICM distribution, the larger the overestimation. 
\item At large distances from the cluster centre the difference for the total mass estimates is negligible for compressed and elongated shapes: at $r_{\mathrm{500}}$ these differences are generally less than $4 \%$.
\item The gas mass fractions are overestimated if the cluster is compressed along the line of sight.
\item If the cluster is elongated along the line of sight  the gas mass
fractions are slighty underestimated, mainly at large radii.
\item We find that projection effects are important when comparing X-ray and lensing mass estimates and that quite large discrepancies between the two methods can at least partially be reconciled if the ICM is elongated along the line of sight. 

\end{itemize} 
Since the core radius along the line of sight is not measurable by means of X-ray observations only, we use SZ measurements to improve the modelling of the ICM distribution with a triaxial model. In fact, assuming a value for the Hubble constant, we are able to estimate the elongation of the ICM along the line of sight and to constrain more precisely total mass and gas mass fraction. In this context SZ maps are very important: the modelling of the SZ temperature decrement with a $\beta$-model does not only allow the comparison of the resulting best fit parameters with those from X-ray analysis, it gives better estimates of the central Compton parameter than single measurements through the cluster centre. This method can clearly also be used if the X-ray emission shows circular isophotes. Consequently, X-ray and SZ maps together, make the investigation of selection effects in X-ray clusters possible.
%
\begin{acknowledgements}
We thank D. Puy for useful discussions. We are grateful to the referee for many useful
suggestions. This work is partially
supported by the Swiss National Science Foundation.
\end{acknowledgements}
%

%

\bibliographystyle{aa}

\end{document}